\documentclass{article}
\usepackage{background}
\backgroundsetup{
  scale=4,
  color=black,
  opacity=0.2,
  angle=45,
  position=current page.center,
  contents={}
}

\usepackage{longtable}

\usepackage{enumitem}
\usepackage{mdframed}
\usepackage{pdfpages}
\usepackage{fullpage}
\usepackage{tabularx,booktabs}
\usepackage{wrapfig}
\usepackage[utf8]{inputenc} 
\usepackage[T1]{fontenc}    
\usepackage{hyperref}       
\usepackage{url}            
\usepackage{booktabs}       
\usepackage{amsfonts}       
\usepackage{nicefrac}       
\usepackage{microtype}      
\usepackage{lipsum}
\usepackage{fancybox}   
\usepackage[export]{adjustbox}
\usepackage{fancyhdr}       
\usepackage{graphicx}       
\graphicspath{{media/}}     
\usepackage{seqsplit}
\usepackage{pgffor} 
\usepackage{ulem}
\usepackage{amssymb}
\usepackage{pifont}
\usepackage{pifont}
\newcommand{\cmark}{\textcolor{green!60!black}{\ding{51}}} 
\newcommand{\xmark}{\textcolor{red}{\ding{55}}}            

\usepackage{tabularx}
\usepackage{enumitem}
\usepackage{listings}
\usepackage{xcolor}

\lstdefinelanguage{json}{
    basicstyle=\ttfamily\small,
    showstringspaces=false,
    breaklines=true,
    morestring=[b]",
    morecomment=[l]{//},
    stringstyle=\color{blue},
    commentstyle=\color{gray},
}

\usepackage{subcaption}
\usepackage{amsthm,amssymb,amsfonts}
\usepackage{tikz,lipsum,lmodern}
\usepackage[most]{tcolorbox}
\definecolor{cream}{RGB}{222,217,201}

\usepackage{xcolor}
\usepackage{listings}

\usepackage{caption}
\usepackage{PRIMEarxiv}
\usepackage{multirow}
\usepackage[utf8]{inputenc} 
\usepackage[T1]{fontenc}    
\usepackage{hyperref}       
\usepackage{url}            
\usepackage{booktabs}       
\usepackage{amsfonts}       
\usepackage{nicefrac}       
\usepackage{microtype}      
\usepackage{lipsum}
\usepackage{fancyhdr}       
\usepackage{graphicx}       
\graphicspath{{media/}}     

\usepackage{tabularx,booktabs}
\usepackage{wrapfig}
\usepackage[utf8]{inputenc} 
\usepackage[T1]{fontenc}    
\usepackage{hyperref}       
\usepackage{url}            
\usepackage{booktabs}       
\usepackage{amsfonts}       
\usepackage{nicefrac}       
\usepackage{microtype}      
\usepackage{lipsum}
\usepackage{fancyhdr}       
\usepackage{graphicx}       
\graphicspath{{media/}}     
\usepackage{seqsplit}
\usepackage{geometry}
\usepackage{caption} 

\usepackage{subcaption}
\usepackage{amsthm,amssymb,amsfonts}
\usepackage{tikz,lipsum,lmodern}
\usepackage[most]{tcolorbox}
\definecolor{cream}{RGB}{222,217,201}

\newtcolorbox{Box1}[2][]{
                lower separated=false,
                colback=white!80!gray,
colframe=black, fonttitle=\bfseries,
colbacktitle=black!50!gray,
coltitle=black,
enhanced,
attach boxed title to top left={xshift=0.5cm,yshift=-2mm},
title=#2,
boxrule=0.5pt,
boxed title style={colframe=black, boxrule=0.5pt},
#1}

\newtcolorbox{Box2}[2][]{
                lower separated=false,
                colback=white!80!white,
colframe=black, fonttitle=\bfseries,
colbacktitle=white!50!white,
coltitle=black,
enhanced,
attach boxed title to top center={yshift=-2mm},
title=#2,
boxrule=0.5pt,
boxed title style={colframe=black, boxrule=0.5pt},
#1,}

\hypersetup{
pdftitle={Autonomous Inorganic Materials Discovery via Multi-Agent Physics-Aware Scientific Reasoning
},
pdfsubject={q-bio.NC, q-bio.QM},
pdfauthor={Markus J. Buehler, MIT, mbuehler@MIT.EDU},
pdfkeywords={Scientific Artificial Intelligence, Multi-agent system, Large Language Model, Materials Design, Scientific Discovery},
}
  
\title{\bf{Autonomous Inorganic Materials Discovery via Multi-Agent Physics-Aware Scientific Reasoning}
\thanks{\textit{\underline{Citation}}: 
\textbf{A. Ghafarollahi, M.J. Buehler. arXiv, DOI:000000/11111., 2025}} 
}

\author{
  Alireza Ghafarollahi \\
  Laboratory for Atomistic and Molecular Mechanics (LAMM)\\Massachusetts Institute of Technology\\ 77 Massachusetts Ave.\\ Cambridge, MA 02139, USA 
   \And
  Markus J. Buehler \\
  Laboratory for Atomistic and Molecular Mechanics (LAMM)  \\
  Center for Computational Science and Engineering\\ Schwarzman College of Computing\\ Massachusetts Institute of Technology\\77 Massachusetts Ave.\\Cambridge, MA 02139, USA\\ \\
  Correspondence: \texttt{mbuehler@MIT.EDU} \\
}

\begin{document}
\maketitle
\begin{abstract}
Conventional machine learning approaches accelerate inorganic materials design via accurate property prediction and targeted material generation, yet they operate as single-shot models limited by the latent knowledge baked into their training data. A central challenge lies in creating an intelligent system capable of autonomously executing the full inorganic materials discovery cycle, from ideation and planning to experimentation and iterative refinement. We introduce SparksMatter, a multi-agent AI model for automated inorganic materials design that addresses user queries by generating ideas, designing and executing experimental workflows, continuously evaluating and refining results, and ultimately proposing candidate materials that meet the target objectives. SparksMatter also critiques and improves its own responses, identifies research gaps and limitations, and suggests rigorous follow-up validation steps, including DFT calculations and experimental synthesis and characterization, embedded in a well-structured final report. The model’s performance is evaluated across case studies in thermoelectrics, semiconductors, and perovskite oxides materials design. The results demonstrate the capacity of SparksMatter to generate novel stable inorganic structures that target the user's needs. Benchmarking against frontier models reveals that SparksMatter consistently achieves higher scores in relevance, novelty, and scientific rigor, with a significant improvement in novelty across multiple real-world design tasks as assessed by a blinded evaluator. These results demonstrate SparksMatter`s unique capacity to generate chemically valid, physically meaningful, and creative inorganic materials hypotheses beyond existing materials knowledge. 
\end{abstract}


\section{Introduction}
The design of novel inorganic materials underpins progress across diverse scientific and engineering domains, from next-generation batteries and catalysts to advanced semiconductors and high-performance structural materials \cite{croguennec2015recent, ning2017bandgap, alberi20182019, zhao2019theory, zhao2020designing}. Historically, materials innovation has relied on empirical exploration, domain intuition, and time-consuming experimental or computational screening. While high-throughput density functional theory (DFT) calculations have accelerated discovery in recent years, the sheer scale and complexity of chemical and structural spaces remain formidable barriers \cite{saal2013materials}, calling for the need for more scalable yet accurate approaches.

Data-driven and machine learning methods have become a transformative force in materials science to accelerate the discovery of promising materials from massive chemical and compositional spaces \cite{guo2021artificial, reiser2022graph, choudhary2022recent, merchant2023scaling}. Models trained on open materials databases \cite{jain2013commentary,kirklin2015open, curtarolo2012aflow} can predict material properties with remarkable accuracy and at speeds far beyond those achievable with traditional first-principles methods \cite{xie2018crystal, park2020developing}. Generative models target inverse materials design by generating novel material structures, unconditionally or conditioned on target properties \cite{xie2021crystal, ren2022invertible, zhao2023physics, zeni2025generative}. AI has also significantly advanced the accurate simulation of inorganic materials through the development of foundational machine-learned force fields \cite{chen2022universal,batatia2022mace, deng2023chgnet, batatia2023foundation, barroso2024open, yang2024mattersim}. Most recently, large language models (LLMs) \cite{wei2022emergent,bubeck2023sparksartificialgeneralintelligence} have marked a paradigm shift in materials science, contributing to various aspects including knowledge extraction and reasoning \cite{buehler2024generative, yang2025learning}, hypothesis generation \cite{buehler2024graphreasoning, ghafarollahi2024sciagents}, materials design \cite{ni2024forcegen, buehler2023melm, buehler2024cephalo} and property prediction \cite{liu2025large}

Despite these advances, existing approaches to inorganic materials design remain fragmented and inadequate for end-to-end autonomous discovery. As shown in Table \ref{tab:models_comparison}, generative models can propose novel structures but lack property evaluation, limiting their practical utility. Surrogate models provide fast predictions but struggle to generalize beyond their training data, especially for unseen properties or compositions. Databases are confined to known compounds and do not support exploration of new materials. More critically, existing tools lack the capacity for reasoning, adaptive planning, and iterative decision-making, rendering them insufficient for autonomous materials discovery. While LLMs introduce new capabilities in reasoning and reflection, their isolated use remains inadequate for the demands of inorganic materials design, which requires physically grounded validation, multi-step workflows, and integration of domain-specific simulations and data.

To overcome these challenges, LLM-driven multi-agent systems have emerged, combining the reasoning capabilities of LLMs with the power of specialized tools \cite{hafner2006toward, jain2016computational, szymanski2023autonomous, tom2024self}. These systems enable the orchestration of specialized LLM agents and support seamless integration with external tools, such as deep learning models or physics-based simulators to solicit physics and enforce domain-specific constraints. Notably, such systems can be designed to be self-improving, continuously augmenting their capabilities by learning from prior results, adapting strategies, and incorporating new knowledge. LLM-based multi-agent frameworks have demonstrated early promise in accelerating scientific discovery across domains including AI \cite{lu2024ai}, chemistry \cite{m2024augmenting},  biomaterials \cite{ghafarollahi2024protagents, ghafarollahi2024sciagents, ghafarollahi2025sparks}, and alloy design \cite{ghafarollahi2025automating}. However, their application to inorganic materials discovery remains largely unexplored. This highlights the need for intelligent, self-improving agents capable of autonomously generating hypotheses across vast chemical and structural spaces, proposing novel candidate materials, predicting relevant properties, and reasoning about synthesizability and experimental feasibility to accelerate the materials discovery process.

In this work, we present SparksMatter, a multi-agent AI framework for inorganic materials design that integrates the reasoning capabilities of large language models (LLMs) with domain-specific tools. SparksMatter is developed to accelerate and automate the materials design process by performing key tasks such as retrieving materials from repositories, generating novel structures with target properties, and predicting material properties. The system is composed of a suite of specialized LLM agents, each responsible for a specific function within the overall workflow. This agentic architecture forms a concrete foundation for SparksMatter to operate as an autonomous AI scientist capable of addressing user-defined queries.
SparksMatter follows a structured ideation–planning–experimentation–reporting pipeline. During the ideation phase, agents collaboratively generate hypotheses to address the posed design challenge. In the planning phase, these hypotheses are translated into actionable research plans. The experimentation phase involves executing these plans through tool use and evaluation. This cycle continues iteratively until the research objectives are met, at which point the system enters the reporting phase to produce a comprehensive summary of findings. 

SparkMatter is designed to emulate scientific thinking where agents engage in reflection, critique, and revision—continually improving their outputs based on newly gathered information. This in-situ reasoning capability is the driving force behind SparksMatter’s transformation from a static inference engine to a dynamic, goal-oriented system capable of handling the complexity of real-world inorganic materials discovery. Furthermore, SparksMatter is designed to be modular and extensible, allowing seamless integration of new tools and workflows. It is envisioned as a general-purpose AI researcher for autonomous inorganic materials design. We demonstrate SparksMatter’s capabilities across a diverse set of materials design tasks and benchmark its performance against state-of-the-art models, including GPT-4 and O3-deep-research.

\begin{table}[h]
\centering
\caption{Comparison of key materials informatics tools across various functional dimensions, including materials design, property prediction, and materials retrieval. SparksMatter integrates all these capabilities into a unified approach, combining the strengths of conventional models with the reasoning abilities of large language models (LLMs).}
\label{tab:models_comparison}
\begin{tabularx}{1.0\textwidth}{lccccc}
\hline
 & Generative AI & DL surrogate & Materials Rep. & LLMs & SparksMatter (Ours)\\
\hline
Design/Generation & \cmark &   \xmark & \xmark & \cmark & \cmark \\
Modeling/Simulation & \xmark &   \xmark & \xmark & \xmark & \cmark \\
Property Prediction & \xmark &   \cmark & \xmark & \cmark & \cmark \\
Database/Repository & \xmark & \xmark & \cmark & \xmark & \cmark \\
Reasoning/Thinking & \xmark &   \xmark & \xmark & \cmark & \cmark\\
\hline
\end{tabularx}
\end{table}

\section{Results and Discussion}
\subsection{Automating inorganic materials design with SparksMatter}
\begin{figure}[ht!]
\centering
    \includegraphics[width=1\textwidth]{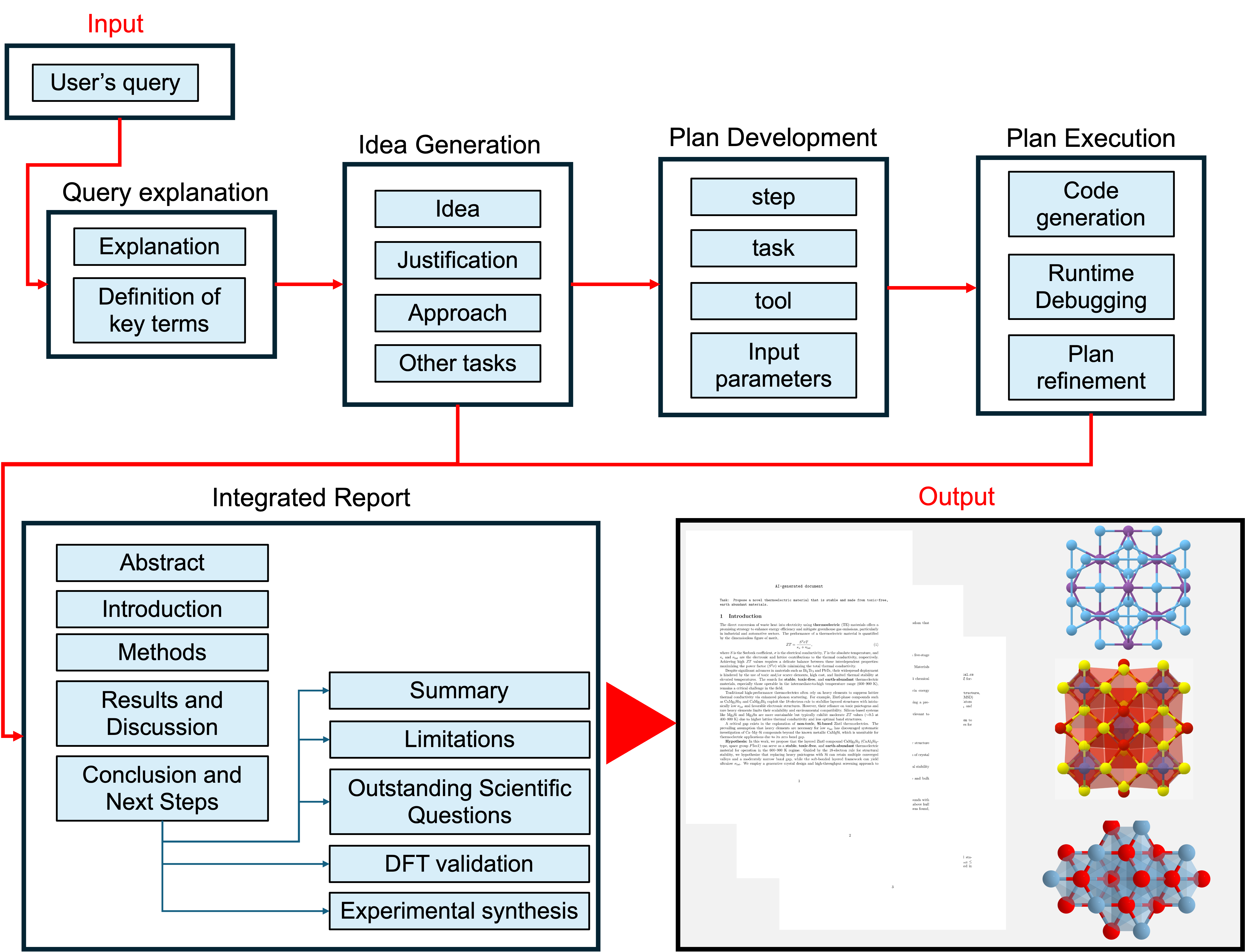}
\caption{Overview of the entire process from initial user's query to the final document, following a structured, yet adaptive strategy where responses are successively examined, refined, and improved. The process begins with the query explanation where the key terms are defined and the query is clarified, setting the stage for the ideation phase. Then, a proposal is developed encompassing critical components such as idea, approach, and other tasks. Then, a detailed structured plan is developed followd by the execution phase where Python codes are generated and executed to follow the plan and create the results. These results, together with idea are then subsequently expanded in the reporting phase to yield a significant amount of additional details, forming a comprehensive document.
}
    \label{fig:workflow}
\end{figure}

The SparksMatter framework operates through an ideation–planning–experimentation–expansion pipeline, as illustrated in Figure~\ref{fig:workflow}. The process begins with a user-defined query that articulates a specific materials design objective, such as discovering a novel, sustainable inorganic compound with targeted mechanical properties.

In the ideation phase, scientist agents interpret the query, define key terms, and frame the scientific context. This lays the groundwork for generating creative hypotheses and formulating a high-level research strategy tailored to the available computational tools. Next, in the planning phase, planner agents translate the high-level strategy into a detailed, executable plan, outlining specific tasks and tool invocations. Each idea and plan is evaluated by designated critics for clarity, accuracy, and completeness before proceeding to the next step.

In the experimentation phase, assistant agents implement the plan: they generate and execute Python code, interact with domain-specific tools, collect intermediate and final results, and store them for final review and reporting. This phase is iterative-agents continuously reflect on the outputs, adapt the plan as necessary, and ensure that all relevant data needed to support the proposed hypothesis is systematically gathered. 

Finally, in the expansion phase, a critic agent reviews the query, idea, plan, and execution results and synthesizes a complete document expanding on various aspects. It assembles the results into a coherent and structured scientific report, addressing the motivation and impact of the task, the methodology and workflow, key findings and their mechanistic interpretation, and limitations of the study, along with recommendations for improvement and future directions.

Empowered by advanced reasoning models like o3, SparksMatter can generate novel ideas and hypotheses, such as previously unconsidered material chemistries that meet sustainability constraints. Through integration with domain-specialized tools, SparksMatter is also capable of designing materials with targeted properties (e.g., band gap, mechanical strength), evaluating their stability, and predicting their performance, thereby ensuring the generated structures are physically reliable. These capabilities distinguish SparksMatter from tool-less LLMs such as o3 and o3-deep-research, promoting both novelty and scientific relevance in materials generation and discovery. Its performance in addressing real-world materials design challenges, evaluated across multiple scientific metrics, is benchmarked against baseline models in Section \ref{sec: benchmark}. The list of tools and functions integrated into the SparksMatter framework is provided in the Materials and Methods section.

\begin{figure}[ht!]
\centering
    \includegraphics[width=1\textwidth]{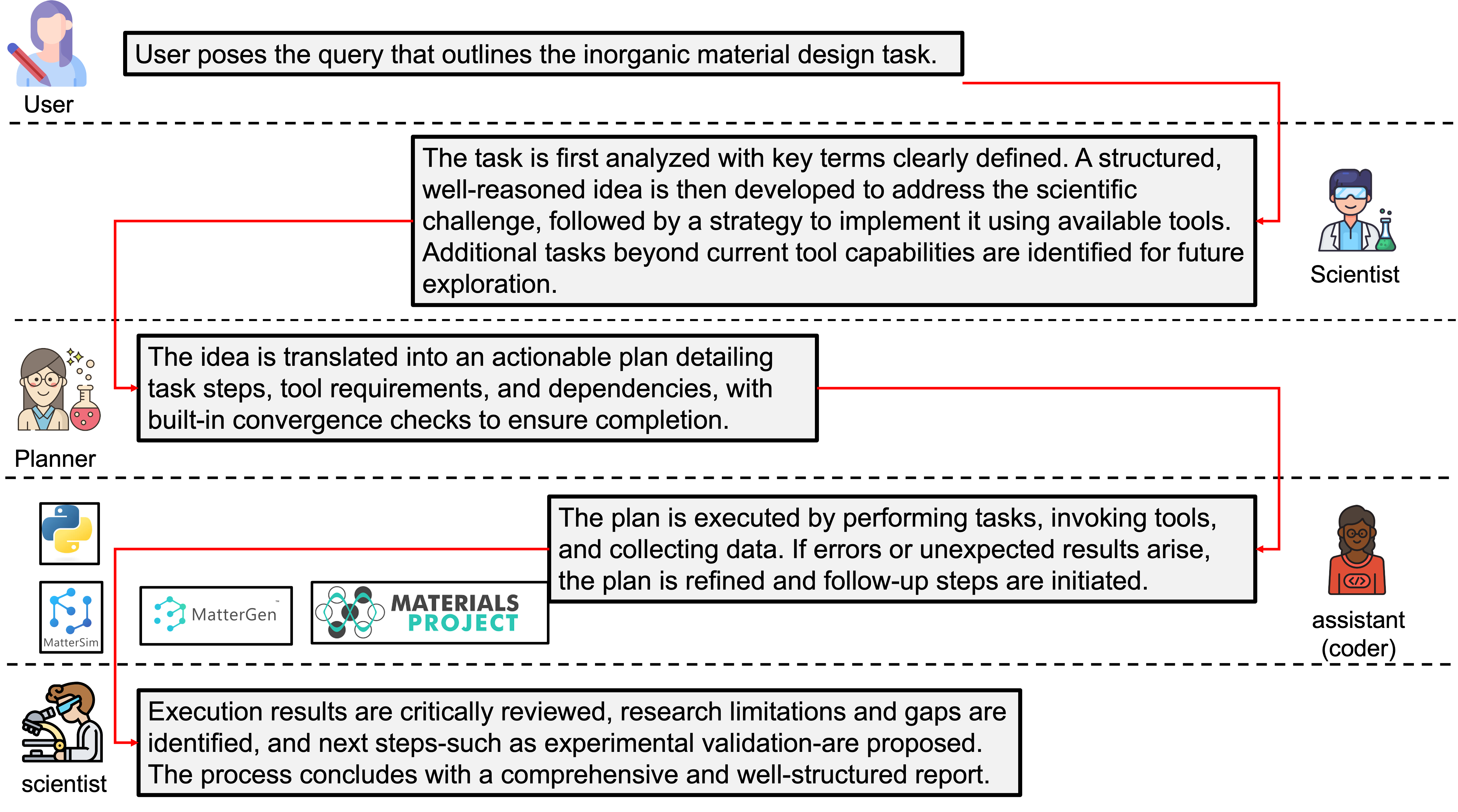}
\caption{Overview of the SparksMatter developed here for automated inorganic material design and analysis. The model comprises a network of specialized AI agents, each responsible for a distinct role; scientist analyzes the user query and clarifies its key terms and proposes a well-reasoned proposal, planner develops a detailed plan to execute the idea with the available tools, and assistant writes Python code to implement the plan, calling computational tools, generate and store results, and refines the plan by proposing follow-up experiments if needed; critic agents review the results, identify limitation and gaps, and provide a well-structured report. These agents operate in a fully self-directed, feedback-driven environment, enabling adaptive decision-making and iterative refinement.}
    \label{fig:overview}
\end{figure}

Figure~\ref{fig:workflow} provides an overview of the workflow conducted by SparksMatter, from the user’s initial query to the generation of the final scientific document and structures. The process begins with a user-defined query that articulates a specific materials design objective. This is followed by a clarification step, where key terms are explained and contextualized.

In the ideation phase, scientist agents are instructed to develop innovative, testable, and scientifically sound ideas to address the posed task. They are prompted to return a structured response comprising several key components: thoughts, idea, justification, approach, and other tasks. Thoughts provide detailed scientific reasoning and theoretical context behind the proposed idea. Idea refers to the core hypothesis or materials design concept. Approach outlines a high-level strategy to test the idea using available tools. Other tasks identify critical steps, such as computational validation or experimental synthesis, that may fall outside the current toolset but are essential for scientific completeness. Next, in the planning phase, a planner agent transforms the high-level idea into a structured sequence of executable steps. Each step includes a clearly defined task, the appropriate tool to be used, and the relevant input parameters.

The execution phase is handled by an assistant agent, which implements the full plan step by step by generating and running Python code. This is where existing materials are retrieved from repositories such as the Materials Project \cite{jain2013commentary, horton2025accelerated}, novel structures are generated using diffusion models like MatterGen \cite{zeni2025generative}, their stability is assessed, and material properties are predicted using deep learning models. After each step, the assistant reflects on the outputs; if unexpected results or issues arise, the plan is refined, and a revised strategy is executed. This feedback-driven, adaptive approach allows for dynamic exploration of the design space, improving both predictive accuracy and procedural efficiency over time. Such adaptability is particularly beneficial in open-ended design challenges, where iteration, optimization, and guided exploration are critical. All generated results, code, and execution notes are stored for full transparency and reproducibility, and made available to the user and the system for the next phase.

In the final documentation phase, agents analyze the original query, the proposed idea, and the collected results. They then refine and enhance the outputs-integrating retrieved data, identifying scientific gaps and limitations, and highlighting important computational and experimental directions that remain unaddressed. The outcome is a well-structured scientific report that presents the motivation, methodology, results, limitations, and suggestions for future work.

SparksMatter thus represents a step toward autonomous scientific reasoning and tool use where complex materials design tasks are navigated by a coordinated ensemble of AI agents capable of reflection, adaptability, and continual improvement. In the sections that follow, we present several real-world applications that demonstrate the efficacy and versatility of this framework.

\subsection{Experiments}
In this section, we present a series of inorganic materials design experiments to demonstrate the effectiveness and versatility of SparksMatter in addressing diverse challenges across inorganic materials design. The full documents generated by SparksMatter for these experiments is provided in SI.

\subsubsection{Task 1: Green and sustainable thermoelectric material design}
As the demand for eco-friendly technologies grows, so does the need for materials that are safe, ethically sourced, and environmentally responsible. Sustainable materials are key to clean energy, low-impact manufacturing, and reduced electronic waste. Here, we show how SparksMatter can identify green thermoelectric materials tailored to specific applications.

\begin{figure}[ht!]
\centering
    \includegraphics[width=1\textwidth]{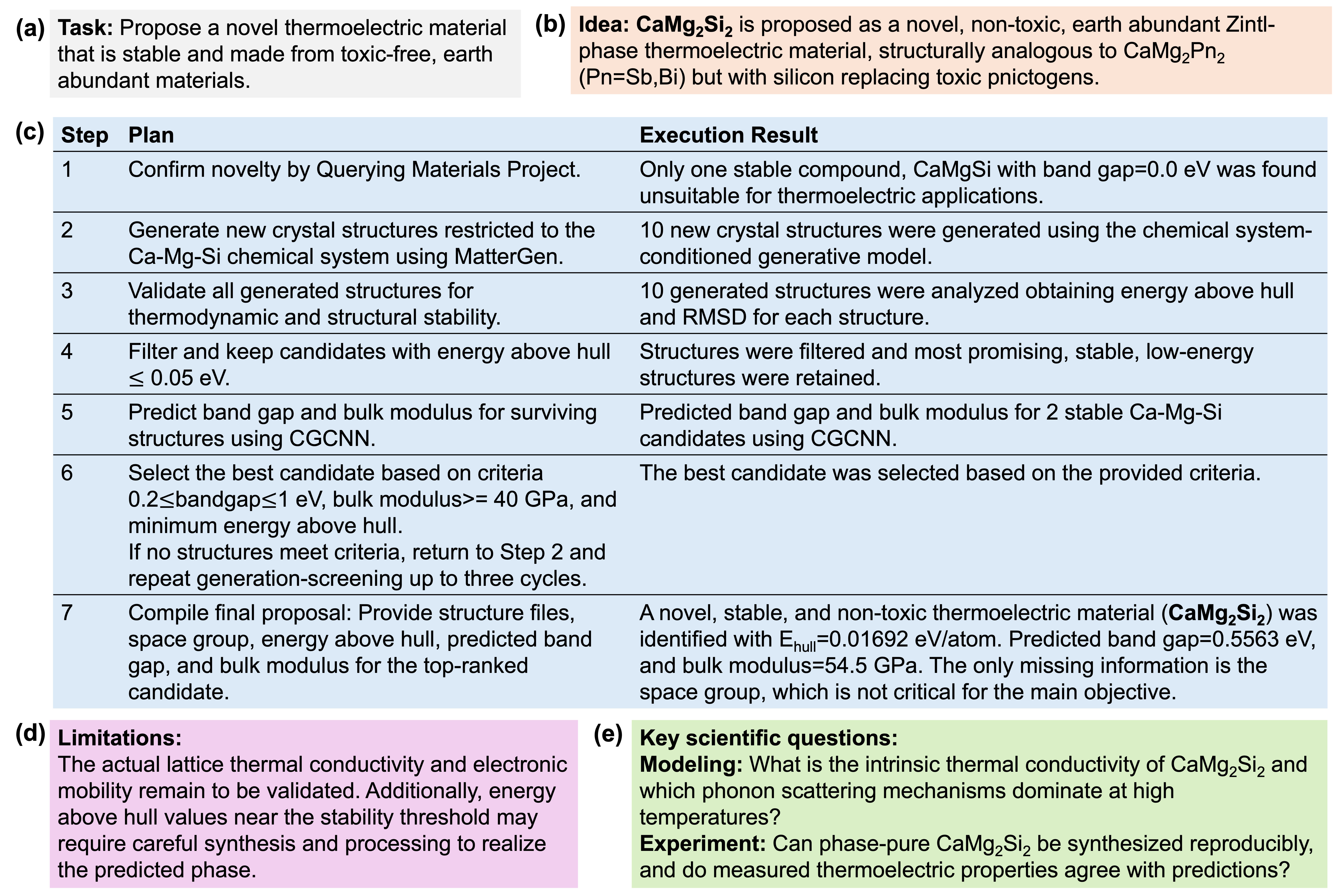}
\caption{\textbf{Overall workflow executed by SparksMatter for Task 1.} \textbf{(a)} User query; \textbf{(b)} Core idea developed by Scientist agents proposing CaMg$_2$Si$_2$ as the novel thermoelectric material candidate; \textbf{(c)} Key plan steps and execution results confirming the initial material hypothesis; \textbf{(d)} Limitations of the research as outlined in the final report; \textbf{(e)} Key modeling and experimental scientific questions identified and proposed for future investigation.}
    \label{fig:task1_results}
\end{figure}

For this example, the user poses the following task: ``Propose a novel thermoelectric material that is stable and made from toxic-free, earth abundant materials.`` The overall workflow of SparksMatter for this task is shown in Figure \ref{fig:task1_results}. The system focuses on the unexplored CaMg$_2$Si$_2$ Zintl phase as the proposed thermoelectric material. The novelty is first confirmed by querying Materials Project for any stable Ca-Mg-Si ternary compounds. Only one stable Ca-Mg-Si compound (CaMgSi, E$_{\mathrm{hull}}$=0, band gap=0.0 eV) which is a metallic compound and not suitable for thermoelectric applications. 

The system then follows an inverse design pipeline, combining generative chemistry-conditioned structure creation with high-throughput thermodynamic screening and machine learning-based property evaluation.

The system begins by conditioning structure generation on the Ca–Mg–Si chemical system. Using its generative model (MatterGen), SparksMatter sampled 10 unique candidate structures. These were then subjected to stability analysis using a pretrained energy model (MatterSim). Structures were filtered based on two criteria: energy above hull $\leq$ 0.05 eV/atom and a binary stability flag indicating success in geometric relaxation. Two structures satisfied these conditions-Ca$_4$Mg$_4$Si$_4$ and CaMg$_2$Si$_2$. For these survivors, SparksMatter predicted key electronic and mechanical properties using a pretrained Crystal Graph Convolutional Neural Network (CGCNN), including band gap and bulk modulus. Among the two candidates, CaMg$_{2}$Si$_2$ emerged as the most promising, exhibiting the lowest energy above hull (0.0169 eV/atom), a moderate band gap (0.5563 eV), and a high bulk modulus (54.49 GPa).

The stability of CaMg$_{2}$Si$_2$ was further rationalized through Zintl chemistry, illustrating SparksMatter’s capacity to integrate domain knowledge beyond its explicit toolset. Although its core tools focus on structure generation, thermodynamic filtering, and property prediction, SparksMatter autonomously inferred that CaMg$_{2}$Si$_2$ satisfies the 18-electron rule-a known criterion for stabilizing Zintl phases. It further identified that the compound likely adopts the CaAl$_2$Si$_2$-type layered structure (space group P–3m1), which is associated with intrinsically low lattice thermal conductivity due to soft interlayer bonding and rattling-like Ca vibrations. Notably, this reasoning challenges the traditional assumption that ultralow $\kappa_{\mathrm{lat}}$ in Zintl thermoelectrics requires heavy elements, highlighting instead the potential of light-element frameworks. The predicted electronic structure, including multiple converged valleys and a moderately narrow band gap, suggests a power factor on par with high-performing compounds such as Mg$_{3}$Sb$_2$. Together with its mechanical robustness and non-toxic, earth-abundant composition, CaMg$_{2}$Si$_2$ emerges as a strong thermoelectric candidate for operation in the 600–900 K range.

SparksMatter also highlighted key limitations, notably the absence of DFT and experimental validation for the proposed structure. To address these gaps and advance CaMg$_2$Si$_2$ toward practical application, SparksMatter outlined a comprehensive follow-up plan that spans both computational validation and experimental realization. First-principles simulations are proposed to confirm the phase's thermodynamic and dynamic stability, including DFT-based structural relaxation, convex hull analysis, and phonon dispersion calculations. For accurate prediction of thermoelectric performance, the system recommends BoltzTraP2 to model electronic transport coefficients and ShengBTE for phonon-mediated lattice thermal conductivity. Dopability will be assessed via defect formation energy calculations under various growth conditions to guide potential n- or p-type doping strategies. For experimental synthesis, SparksMatter suggests solid-state reaction routes such as spark plasma sintering of Ca, Mg, and Si powders, followed by phase confirmation using XRD and microstructural analysis with SEM and EDS. Transport properties will be evaluated from room temperature to 900 K using Seebeck coefficient, resistivity, and thermal conductivity measurements. Long-term stability will be probed through thermal cycling and oxidation resistance studies.

The full document generated by SparksMatter for this example is provided in \ref{sec: task1}  of SI. These results demonstrate SparksMatter’s capability to autonomously propose chemically valid, thermodynamically plausible, and application-relevant material candidates. The framework not only generates novel hypotheses but also constructs end-to-end workflows to guide experimental realization, thereby enabling closed-loop, data-driven discovery in sustainable energy materials.

\subsubsection{Task 2: Inorganic soft semiconductors design}
Next, SparksMatter is queried with the task: ``Propose novel semiconductors alternative to organic materials that are mechanically soft (bulk modulus < 30 GPa) and thermodynamically stable.``
This query addresses a critical challenge in materials design for flexible electronics, where mechanical softness and environmental stability are essential. Organic semiconductors provide the necessary flexibility but suffer from limited carrier mobility, thermal instability, and degradation under ambient conditions. In contrast, conventional inorganic semiconductors are typically too stiff for applications requiring mechanical compliance. Bridging this materials gap-by identifying soft, stable, and purely inorganic semiconductors-could unlock a new generation of durable, high-performance components for wearable and bendable devices.

\begin{figure}[ht!]
\centering
    \includegraphics[width=1\textwidth]{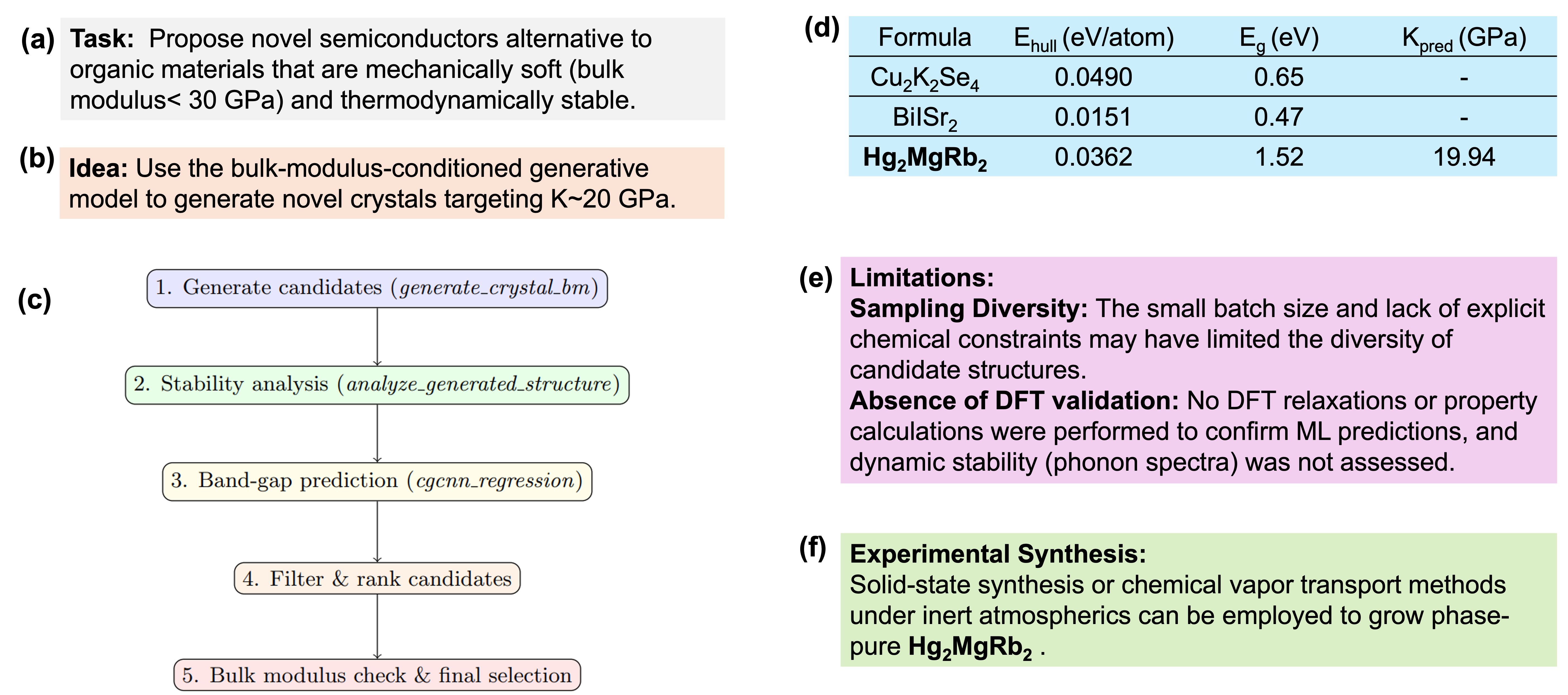}
\caption{\textbf{Overall workflow conducted by SparksMatter for Task 2.} \textbf{(a)} User-defined task; \textbf{(b)} Core idea proposed by Scientist agents; \textbf{(c)} Computational workflow diagram generated by SparksMatter and included in the final document; \textbf{(d)} Inorganic material candidates and their predicted properties. Hg$_2$Mg$_2$Rb$_2$ was selected as the final candidate for a soft inorganic semiconductor; \textbf{(e)} Research limitations as documented in the final report; \textbf{(f)} Recommendations for experimental synthesis of the selected candidate.}
    \label{fig:task2_results}
\end{figure}

To address this task, SparksMatter activates an inverse-design workflow that combines property-conditioned generative modeling with multi-stage screening and prediction. The overall workflow conducted by  SparksMatter is shown in Figure \ref{fig:task2_results}. The process begins with the generation of candidate crystal structures using a property-conditioned model (MatterGen) targeting a bulk modulus near 20 GPa. The generated candidates are then screened for thermodynamic stability using energy-above-hull calculations, and their electronic band gaps are predicted using CGCNN. Candidates satisfying all design criteria-mechanical softness (K < 30 GPa), semiconducting band gap (0.8-2.0 eV), and low formation energy-are retained for further evaluation.

From a pool of eight generated structures, SparksMatter identifies Hg$_2$MgRb$_2$ as a purely inorganic compound meeting all constraints: a predicted bulk modulus of 19.94 GPa, a band gap of 1.52 eV, and energy above hull of 0.036 eV/atom. 

SparksMatter also analyzes the underlying mechanisms governing the proposed material  behavior, offering insights into both structural and electronic properties. It attributes the mechanical softness of Hg$_2$MgRb$_2$ to its layered structure and large Rb ions, which weaken interlayer bonding and reduce lattice stiffness. The inclusion of heavy cations like Hg and Mg further lowers the bulk modulus by attenuating bond force constants. The model also explains the 1.52 eV band gap as arising from hybridization between Hg 6s, Rb 5s, and anion states, yielding an electronic profile similar to hybrid perovskites but with enhanced stability due to the absence of organic components. Together, these results highlight SparksMatter’s expert-level capacity to infer and generalize structure-property relationships.

In addition to candidate identification, SparksMatter provides a comprehensive roadmap for follow-up validation. The proposed next steps include first-principles calculations of elastic tensors and phonon spectra, finite-temperature simulations to assess dynamic stability, and defect analysis to evaluate dopability and charge transport. Experimental synthesis routes (e.g., solid-state or vapor-phase growth), thin-film processing strategies, and environmental assessments are also highlighted and recommended by the model. 

The full document created by SparksMatter for this experiment is provided in \ref{sec: task2} of SI. This case study exemplifies SparksMatter’s integrated approach to autonomous materials discovery—merging generative modeling, machine-learned property prediction, and LLM-based scientific reasoning into a coherent, expert-like workflow. Beyond identifying promising candidates, the system interprets structure–property relationships, proposes mechanistic explanations, and outlines rigorous computational and experimental validation strategies. Its ability to connect structural features to macroscopic behavior and anticipate viable synthesis pathways reflects a level of scientific intuition typically reserved for human experts.

\subsubsection{Task 3: A toxic-free perovskite oxide material}
In this example, SparksMatter is queried with the task:
``Identify a toxic-free perovskite oxide material like PbTiO$_3$.``
This task addresses replacing PbTiO$_3$-a widely used ferroelectric perovskite-with a compositionally safe, environmentally benign alternative that preserves its superior piezoelectric and ferroelectric properties. 

SparksMatter addresses the challenge by proposing a data-driven workflow to identify and validate environmentally benign ABO$_3$ perovskites with comparable functionality. Focusing on the promising lead-free candidate Na$_{0.5}$K$_{0.5}$NbO$_3$ (KNN), SparksMatter leverages the Materials Project to retrieve all known K-Na-Nb-O crystal structures and filter them based on thermodynamic stability (energy above hull $\leq$ 0.1 eV). The candidates are then passed through a Crystal Graph Convolutional Neural Network (CGCNN) to predict their electronic and mechanical properties. Finally, SparksMatter benchmarks the predicted values-band gap, bulk modulus, and formation energy-against reference values for PbTiO$_3$ to assess viability.

Through this autonomous workflow, SparksMatter identifies two viable candidates with the formula KNaNb$_2$O$_6$. Both structures exhibit low energy above the convex hull (<0.03 eV/atom), band gaps in the range of 2.41-2.44 eV, and bulk moduli near 98 GPa. These predictions indicate that KNaNb$_2$O$_6$ approximates the key functional characteristics of PbTiO$_{3}$ while avoiding the use of toxic elements. Notably, although the tools invoked by SparksMatter do not explicitly include polarization or phase-transition models, the system reasons beyond its toolset-drawing on structural motifs, valence electron configurations, and known chemistries-to suggest KNaNb$_22$O$_6$ as a promising ferroelectric candidate. This demonstrates SparksMatter’s capacity to extend its inference beyond direct property predictions and to emulate expert-like materials reasoning.

In addition to identifying candidates, SparksMatter outlines a forward-looking plan for computational and experimental validation. This includes proposals for Berry-phase calculations of spontaneous polarization, phonon-based Curie temperature estimation, defect modeling, and domain engineering. It also recommends experimental synthesis via solid-state or sol–gel routes, along with microstructural and functional testing across temperature and field ranges. This response illustrates SparksMatter’s end-to-end capability, not only to generate and evaluate candidates, but to guide actionable next steps toward realizing sustainable, high-performance, lead-free perovskite materials.

\subsection{Benchmark}\label{sec: benchmark}
To evaluate the performance of the SparksMatter framework, we benchmarked its responses against three baseline reasoning models developed by OpenAI: o3, o3-deep-research, and o4-mini-deep-research. Each model was instructed to act as an expert chemist with access to \textit{internet browsing} but no integration with external scientific tools such as diffusion models. All models were presented with the same set of queries corresponding to Tasks 1, 2, and 3. Their responses were collected and then submitted to a separate evaluator LLM (GPT-4.1) along with the final document generated by SparksMatter, which was tasked with critically assessing each submission. The evaluator highlighted the strengths and weaknesses of each model and scored each response on a scale of 1 to 5 across four key metrics: Relevance (how well the response addresses the task), Scientific Soundness (validity of methods, data, and conclusions), Novelty (originality of ideas or approaches), and Depth and Rigor (quality and completeness of analysis and reasoning).

\begin{figure}[ht!]
\centering
    \includegraphics[width=1\textwidth]{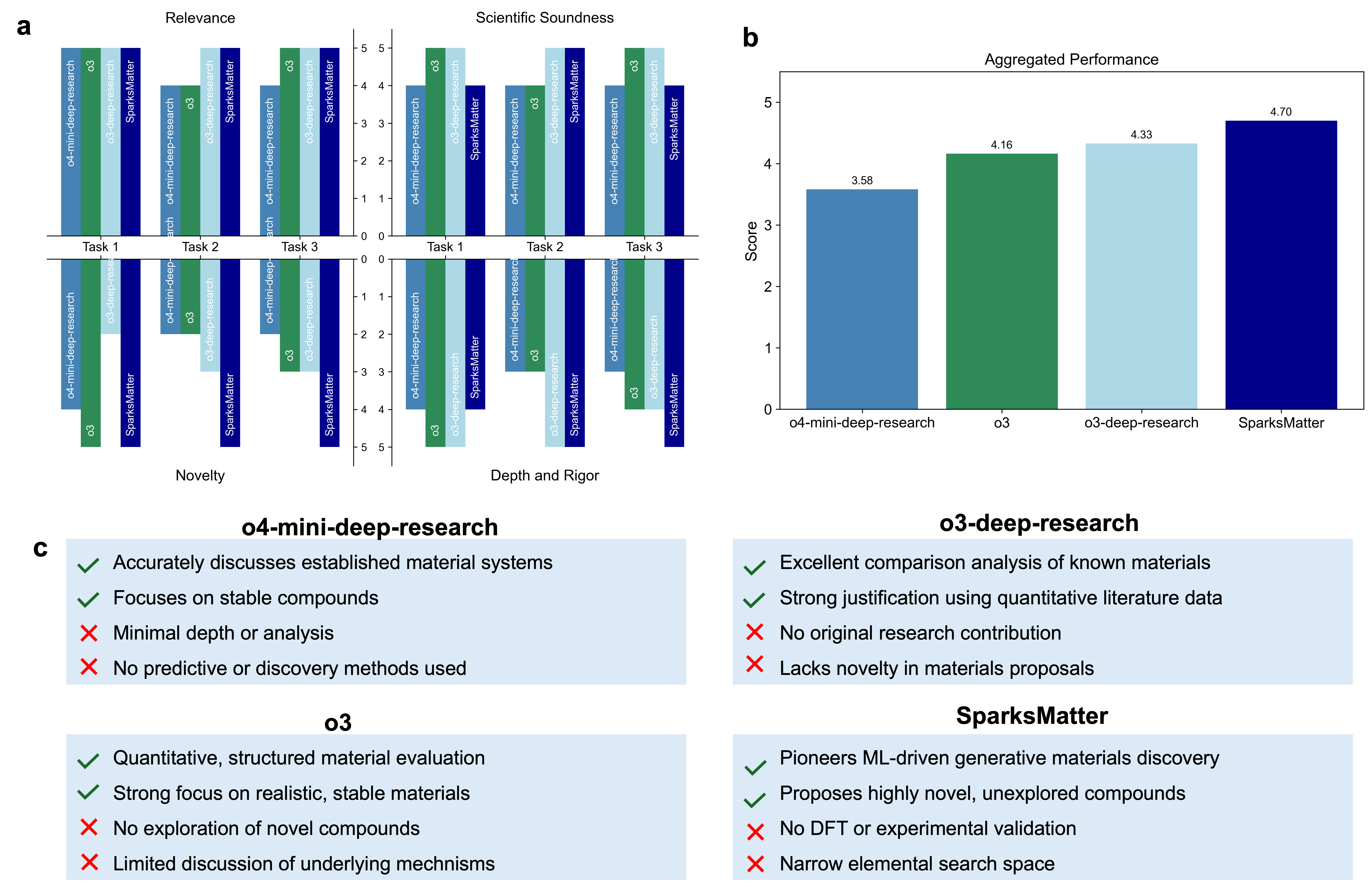}
\caption{\textbf{Comparative performance of SparksMatter and OpenAI reasoning models (o4-mini-deep-research, o3, and o3-deep-research).} 
\textbf{a} Per-task performance comparison based on evaluations by GPT-4.1, using the criteria of relevance, scientific soundness, novelty, and depth and rigor. 
\textbf{b} Aggregated performance scores for each model across all tasks. 
\textbf{c} Strengths and weaknesses of each model as identified by the evaluator.}

    \label{fig:benchmark}
\end{figure}

The full evaluation of SparksMatter and OpenAI model responses for Tasks 1, 2, and 3 is presented in Sections~\ref{sec: task1_evaluation}, \ref{sec: task2_evaluation}, and \ref{sec: task3_evaluation} of the SI, respectively. The evaluation scores are shown in Figure~\ref{fig:benchmark}(a) with aggregated performance provided in Figure~\ref{fig:benchmark}(b). Evidently, SparksMatter consistently outperforms the baseline models across most metrics—particularly in Novelty and Depth and Rigor. In contrast, the baseline models performed poorly on Novelty, often focusing on well-established materials without original calculations or synthesis. This highlights the importance of combining generative models with external tools to enable creative, data-driven exploration in inorganic materials design, as demonstrated by SparksMatter.

While SparksMatter demonstrates strong performance across most evaluation criteria, it shows a modest limitation in Scientific Soundness. This is primarily due to the absence of direct validation for the proposed materials using first-principles methods such as DFT or supporting experimental evidence. Additionally, some key properties, like lattice thermal conductivity, were not explicitly calculated. However, it is worth mentioning that these gaps were explicitly recognized and documented by SparksMatter as recommendations for future development, as highlighted in Figures \ref{fig:task1_results}(d) and \ref{fig:task2_results}(e). Moreover, such limitations can be effectively addressed by integrating first-principles simulators or experimental feasibility predictors as tools. We leave these enhancements to future work, where they can further strengthen SparksMatter’s scientific rigor without altering its core framework.

\section{Conclusion}\label{sec:conclusion}
In this work, we introduced SparksMatter, an LLM-driven multi-agent framework designed to automate the full cycle of inorganic materials discovery. By combining the reasoning, planning, and coding capabilities of large language models with a suite of specialized scientific tools, SparksMatter enables an integrated, closed-loop platform for the autonomous generation, evaluation, and refinement of novel inorganic compounds. Through the coordinated operation of expert agents, the system can propose candidate materials, predict their properties, and iteratively optimize its search strategies-all with minimal human input.

One of the key advantages of SparksMatter lies in its ability to promote scientific novelty by leveraging generative models that extend beyond known materials space. This addresses a central limitation in tool-less reasoning models such as o3-deep-research, which primarily perform knowledge synthesis without producing original hypotheses or exploring uncharted chemical systems.

A central innovation of the framework is its modularity and adaptability. New agents and tools can be seamlessly integrated to expand its capabilities across various areas of computational materials science. This flexibility is particularly important for overcoming current limitations in SparksMatter, such as the lack of essential property predictions (e.g., lattice thermal conductivity in Task 1), by incorporating additional simulation engines or first-principles methods. Moreover, SparksMatter can be extended with fine-tuned LLMs specifically tailored to processing pathways, synthesizability estimation, and experimental constraints, ensuring that the proposed materials are not only promising in theory but also viable in practice.

Overall, SparksMatter supports fully autonomous, interpretable inorganic materials design and acts as a virtual AI research assistant, lowering entry barriers for non-experts and enhancing productivity for domain specialists. Its structured, reproducible, and adaptable workflow positions it as a valuable tool for accelerating innovation in the materials community.

Looking ahead, SparksMatter provides a robust foundation for the next generation of autonomous scientific discovery. Future work will focus on integrating experimental feedback, advancing synthesizability and processing-awareness, and deepening its connection to first-principles validation. By embedding domain-specific expertise and harnessing reinforcement learning, SparksMatter aspires to further accelerate, democratize, and scale the discovery of sustainable and high-performance materials.

\section{Materials and Methods}
\subsection{Agent Design} 
AI agents are implemented using the GPT-4 family of LLMs \cite{achiam2023gpt}. The agentic workflows that support the ideation and experimentation modules are built using AG2 \cite{AG2_2024}, an open-source framework for agent-based AI systems, based on the \texttt{ConversableAgent} class. Critic agents responsible for final evaluation and documentation are instantiated using a custom wrapper function, \texttt{get\_response\_from\_LLM}, which interfaces with the OpenAI API.

Each agent is initialized with a \texttt{system\_message} parameter that defines its role within the system and expected response format. These system messages are composed using detailed prompts, which may include one or more runtime placeholders dynamically updated each time the agent is invoked. The full prompts defining each agent’s \texttt{system\_message} are available in the SparksMatter codebase. 

\subsection{Tools}
All computational tools are implemented as Python functions and stored in the \texttt{functions\_SParksMatter.py}
module. These functions are created outside the agent environment and thus are not inherently known to the SparksMatter agents. The agents are informed by these tools via a detailed description detailing the tool name, full functionality description, input parameters, and output format. Each tool description is provided in the tool's docstring. The execution agent (coder agant) is specifically nstructed to import the relevant functions from the \texttt{functions\_SparksMatter} module to be able to use it in the code to generate the desired results.  

The following tools are implemented in SparksMatter

\subsubsection{Materials Database}
We used the Materials Project \cite{jain2013commentary} as the primary database for existing materials retrieval, accessed via the Materials Project API. Materials and associated metadata were retrieved based on model-specified filter criteria, and the corresponding structures were downloaded in CIF format for downstream applications.

\subsubsection{Generative Material Design}
We employed MatterGen \cite{zeni2025generative}, a generative model for inorganic materials design, for inorganic materials design. Four modes of generation were implemented: (a) unconditional generation, which produces random inorganic materials without constraints; (b) band gap-conditioned generation, targeting materials with specified electronic properties; (c) bulk modulus-conditioned generation, focused on mechanical stiffness; and (d) chemical system-conditioned generation, which generates structures based on a specified chemical system. Each conditional mode guides the generative process toward desired properties or compositions, enabling targeted exploration of materials space.

\subsubsection{Thermodynamic Stability Analysis}
We assessed the thermodynamic stability of the generated structures using the evaluation module provided in \cite{MatterGen_GitHub}. This module leverages MatterSim \cite{yang2024mattersim}, a machine-learned interatomic force field, to perform structure relaxation and construct the convex hull of formation energies.

\subsubsection{Deep learning model for materials property prediction}
We used Crystal Graph Convolutional Neural Networks (CGCNN) \cite{xie2018crystal} as a deep-learning surrogate model for rapid prediction of materials properties, including formation energy, band gap, bulk modulus, and shear modulus.

\subsection*{Conflict of interest}
The author declares no conflict of interest.

\subsection*{Data and code availability}
All data and codes are available on GitHub at \url{https://github.com/lamm-mit/SparksMatter}. 

\subsection*{Supplementary Materials}
Additional materials are provided as Supplementary Materials.

\section*{Acknowledgments}
We acknowledge support from MIT’s Generative AI Initiative. AG gratefully acknowledges the financial support from the Swiss National Science Foundation (project \#P500PT\_214448).

\bibliographystyle{naturemag}
\bibliography{library}

\newpage
\appendix

\pagestyle{empty} 

\renewcommand{\thefigure}{S\arabic{figure}}
\setcounter{figure}{0} 
\renewcommand{\thetable}{S\arabic{table}}
\setcounter{table}{0} 

\clearpage
\begin{center}
\LARGE\bfseries \section*{Supplementary Materials}
\vspace{2cm}

\LARGE\bfseries 
\vspace{1cm}
SparksMatter: Autonomous Inorganic Materials Discovery via Multi-Agent Physics-Aware Scientific Reasoning
\end{center}
\begin{center}

Alireza Ghafarollahi and Markus J. Buehler

\vspace{1cm}
\noindent \textbf{Correspondence:} \texttt{mbuehler@MIT.EDU}
\end{center}

\renewcommand{\thesection}{S\arabic{section}}

\clearpage

\section{document generated by SparksMatter for Task 1}\label{sec: task1}

\backgroundsetup{contents=AI-generated document for Task 1}  

\foreach \pagenum in {1,...,7} {  
  \begin{center}

    \setlength\fboxsep{0pt}  
    
      \includegraphics[width=1\textwidth,page=\pagenum]{./task_1}
    
  \end{center}
  \newpage  
}

\section{document generated by SparksMatter for Taks 2}\label{sec: task2}

\backgroundsetup{contents=AI-generated document for Task 2}

\foreach \pagenum in {1,...,7} {  
  \begin{center}

    \setlength\fboxsep{2pt}  
    
      \includegraphics[width=1\textwidth,page=\pagenum]{./task_3.pdf}
    
  \end{center}
  \newpage  
}

\section{document generated by SparksMatter for Taks 3}\label{sec: task3}

\backgroundsetup{contents=AI-generated document for Task 3}

\foreach \pagenum in {1,...,7} {  
  \begin{center}

    \setlength\fboxsep{2pt}  
    
      \includegraphics[width=1\textwidth,page=\pagenum]{./task_2.pdf}
    
  \end{center}
  \newpage  
}

\section{Evaluation of Task 1 responses from SparksMatter and OpenAI reasoning models, as assessed by GPT-4 evaluator.}\label{sec: task1_evaluation}

\backgroundsetup{contents={}}

\foreach \pagenum in {1,...,4} {  
  \begin{center}
    \setlength\fboxsep{2pt}  
      \includegraphics[width=1\textwidth,page=\pagenum]{./Task1_benchmark.pdf}
  \end{center}
  \newpage  
}

\section{Evaluation of Task 2 responses from SparksMatter and OpenAI reasoning models, as assessed by GPT-4 evaluator.}\label{sec: task2_evaluation}

\backgroundsetup{contents={}}

\foreach \pagenum in {1,...,4} {  
  \begin{center}
    \setlength\fboxsep{2pt}  
      \includegraphics[width=1\textwidth,page=\pagenum]{./Task2_benchmark.pdf}
  \end{center}
  \newpage  
}

\section{Evaluation of Task 3 responses from SparksMatter and OpenAI reasoning models, as assessed by GPT-4 evaluator.}\label{sec: task3_evaluation}

\backgroundsetup{contents={}}

\foreach \pagenum in {1,...,3} {  
  \begin{center}
    \setlength\fboxsep{2pt}  
      \includegraphics[width=1\textwidth,page=\pagenum]{./Task3_benchmark.pdf}
  \end{center}
  \newpage  
}

\end{document}